\pdfoutput=1
\documentclass[
	affil-it, 
	auth-lg, 
	british, 
	compression, 
	contents, 
	custom-packages={}, 
	references={bigone}, 
]{lib/preprint}

\usepackage{cleveref}
\allowdisplaybreaks

\begin{document}

	\hypersetup{
		pdftitle = {Integrability from Homotopy Algebras},
		pdfauthor = {Luigi Alfonsi,Leron Borsten,Mehran Jalali Farahani,Hyungrok Kim,Martin Wolf,Charles Young},
		pdfkeywords = {}
	}

	\date{\today}

	\email{l.alfonsi@herts.ac.uk,l.borsten@herts.ac.uk,m.jalalifarahani@surrey.ac .uk,h.kim2@herts.ac.uk,m.wolf@surrey.ac.uk,c.young8@herts.ac.uk}

	\preprint{DMUS--MP--26/01}

	\title{Integrability from Homotopy Algebras}

	\author[a]{Luigi~Alfonsi\,\orcidlink{0000-0001-5231-2354}\,}
	\author[a]{Leron~Borsten\,\orcidlink{0000-0001-9008-7725}\,}
	\author[b]{Mehran~Jalali~Farahani\,\orcidlink{0009-0002-8282-9316}\,}
	\author[a]{Hyungrok~Kim\,\orcidlink{0000-0001-7909-4510}\,}
	\author[b]{Martin~Wolf\,\orcidlink{0009-0002-8192-3124}\,}
	\author[a]{Charles A.~S.~Young\,\orcidlink{0000-0002-7490-1122}\,}

	\affil[a]{Centre for Mathematics and Theoretical Physics,\\Department of Physics, Astronomy, and Mathematics,\\University of Hertfordshire, Hatfield AL10 9AB, United Kingdom}
	\affil[b]{School of Mathematics and Physics,\\ University of Surrey, Guildford GU2 7XH, United Kingdom}

	\abstract{Homotopy algebraic methods have become increasingly influential in the study of field theories. We consider semi-holomorphic Chern--Simons theory and its relation with the principal chiral model. The semi-classical equivalence of the two theories is expected on general grounds, as the principal chiral model arises by eliminating trivial pairs and gauge fixing. We construct this equivalence explicitly as an off-shell, cyclic quasi-isomorphism between the cyclic $L_\infty$-algebras governing the two theories, and show that it directly encodes the Lax connection. This provides a concrete example for studying integrability of a two-dimensional system through the homotopy algebraic lens.}

	\acknowledgements{We thank Branislav Jur{\v c}o and Christian Saemann for fruitful discussions.}

	\declarations{
		\textbf{Funding.}
		MJF acknowledges support by the STFC Consolidated Grant ST/X000656/1.\\[5pt]
		\textbf{Conflict of interest.}
		The authors have no relevant financial or non-financial interests to disclose.\\[5pt]
		\textbf{Data statement.}
		No additional research data beyond the data presented and cited in this work are needed to validate the research findings in this work.\\[5pt]
		\textbf{Licence statement.}
		For the purpose of open access, the authors have applied a Creative Commons Attribution (CC-BY) license to any author-accepted manuscript version arising.
	}

	\begin{body}

		\section{Introduction and conclusions}

		\paragraph{Background.}
		Homotopy algebras have emerged as a natural algebraic framework for perturbative (quantum) field theory formulated in the Batalin--Vilkovisky formalism. In this approach, the field content, gauge symmetries, equations of motion, and Noether identities of a theory are encoded in an $L_\infty$-algebra through its homotopy Lie brackets on a graded vector space of fields, ghosts, antifields, and antifield ghosts, respectively. The classical Batalin--Vilkovisky action is then given by the corresponding homotopy Maurer--Cartan action. See e.g.~\cite{Hohm:2017pnh,Jurco:2018sby} for detailed accounts of this perspective.

		It thus follows that many central aspects of (quantum) field theory may be articulated in terms of homotopy algebra: scattering amplitudes, generalised amplitudes, and Witten diagrams arise from minimal models of the underlying (relative) $L_\infty$-algebras~\cite{Nutzi:2018vkl,Macrelli:2019afx,Arvanitakis:2019ald,Jurco:2019yfd,Lopez-Arcos:2019hvg,Jurco:2020yyu,Saemann:2020oyz,Gomez:2020vat,Bonezzi:2023xhn,Borsten:2024dvq}; colour--kinematics duality, in a variety of Yang--Mills-matter theories, non-linear sigma models, and Chern--Simons-matter (and M2-brane) theories, is a manifestation of a kinematic  homotopy algebra~\cite{Reiterer:2019dys,Borsten:2020zgj,Borsten:2021gyl,Borsten:2021hua,Ben-Shahar:2021zww,Borsten:2022vtg,Borsten:2022srni,Borsten:2022ouu,Borsten:2023ned,Borsten:2023reb,Borsten:2023paw,Bonezzi:2023pox,Bonezzi:2024dlv,Armstrong-Williams:2024icu,Bonezzi:2024fhd}; the equivalence of twistor actions for gauge and gravity theories may be established through $L_\infty$-quasi-isomorphisms \cite{Borsten:2023paw,Borsten:2024cfx}; equivalences and dualities may be understood as  spans of $L_\infty$-algebras~\cite{JalaliFarahani:2023sfq}; topological/holomorphic twists, supersymmetric localisation, spontaneous symmetry breaking, and anomalies are all realised uniformly as twists of (quantum) $L_\infty$-algebras~\cite{Elliott:2020uwn,Borsten:2025hrn,Borsten:2025dyv}; generalised symmetries and their associated phases admit homotopy algebraic  descriptions through Symmetry Topological Field Theories (SymTFTs) constructed via the Alexandrov--Kontsevich--Schwarz--Zaboronsky formalism~\cite{Borsten:2025pvq}; aspects of holography can be phrased in terms of homotopy transfer~\cite{Chiaffrino:2023wxk}, relative $L_\infty$-algebras~\cite{Alfonsi:2024utl}, and Koszul duality~\cite{Paquette:2021cij}; perturbative black hole solutions in (super)gravity can be understood through homotopy transfer~\cite{Gutowski:2025lzi}; non-commutative field theories via homotopy algebras~\cite{DimitrijevicCiric:2021jea,Nguyen:2021rsa,Giotopoulos:2021ieg,DimitrijevicCiric:2023hua,Bogdanovic:2024jnf}; fluid dynamics as homotopy algebra~\cite{Escudero:2022zdz,Napper:2023jky} and effective hydrodynamic limits of (quantum) field theories as cospans of differential graded manifolds~\cite{Jonsson:2026nmd}. This list is certainly not exhaustive, but gives an impression of the range of topics that have thus far been considered in homotopy algebraic terms.

		Very recent work~\cite{Benini:2026pwa} (see also~\cite{Benini:2020skc}), which appeared whilst the present work was in preparation, has pushed this principle  yet further, applying it to the study of integrable systems arising from four-dimensional semi-holomorphic Chern--Simons theory~\cite{Costello:2017dso,Costello:2018gyb,Costello:2019tri}. In particular, the integrability structures of two-dimensional theories associated with the Chern--Simons theory are shown to descend from homotopy transfer.\footnote{See~\cite{Bittleston:2026tdr} for an off-shell discussion between four-dimensional semi-holomorphic Chern--Simons theory and the Hitchin system in two dimensions.}

		\paragraph{Summary of results.}
		In the present work, we make explicit the perturbative equivalence between four-dimensional semi-holomorphic Chern--Simons theory, defined with a specific meromorphic one-form measure, and the two-dimensional principal chiral model in the language of $L_\infty$-algebras. Whilst such an equivalence is expected on general grounds (see \cref{sec:quasi-iso}), we construct an explicit, off-shell, cyclic quasi-isomorphism between the cyclic $L_\infty$-algebras governing the two theories. This provides a systematic and explicit derivation of their semi-classical equivalence.\footnote{See e.g.~\cite[Section 3.4]{Borsten:2021gyl} for a discussion on the different equivalences of field theories in this context.} It also directly provides the Lax connection of the principal chiral model which appears to be encoded in the quasi-isomorphism.

		In \cref{sec:PCMfromSHCS} we briefly review the relationship between semi-holomorphic Chern--Simons theory and the principal chiral model as it is understood in the literature. With this background, we turn to the cyclic $L_\infty$-algebra of Batalin--Vilkovisky semi-holomorphic Chern--Simons theory in \cref{ssec:LinftyStructureCS}. In particular, we give the explicit $L_\infty$-brackets and cyclic structure of the Batalin--Vilkovisky theory and, in particular, the boundary/pole conditions that ensure cyclicity holds. The corresponding analysis for the principal chiral model is then given in \cref{sec:PCM}, following~\cite{JalaliFarahani:2023sfq}. With these two cyclic $L_\infty$-algebra formulations at hand, we are then able to construct an explicit quasi-isomorphism between them, establishing their equivalence in \cref{sec:quasi-iso}.

		\paragraph{Conclusions, outlook, and the relative $L_\infty$-algebra perspective.}
		Firstly, it would be interesting to see if the quasi-isomorphism presented in the present work is given by homotopy transfer as in~\cite{Benini:2026pwa}. If this is not the case, then one ought to establish a corresponding span of $L_\infty$-algebras as in~\cite{JalaliFarahani:2023sfq}, where the various quasi-isomorphims arise from homotopy transfer. Furthermore, given our example, there are many evident generalisations that are almost immediate. To mention but a few: establishing quasi-isomorphisms amongst the two-/four-dimensional duals of the variety of generalised four-dimensional Chern--Simons theories parametrised by meromorphic one-forms \cite{Ashwinkumar:2023zbu,Levine:2023wvt,Schmidtt:2023slc,Schmidtt:2025zsb,Fukushima:2025tlj,Cole:2025zmq}; constructing the Batalin--Vilkovisky theory and $L_\infty$-algebra of five-dimensional Chern--Simons theory~\cite{Costello:2016nkh,Ashwinkumar:2024vys}, and examining the proposed (M-theory) dualities and  equivalences amongst lower-dimensional theories;  encoding as quasi-isomorphisms and $L_\infty$-spans the web of relations amongst four-dimensional Chern--Simons theory, the anti-self-dual Yang--Mills equations, six-dimensional holomorphic Chern--Simons theory on twistor space, and the principal chiral model, its integrable deformations, coset models, and non-Abelian T-duals~\cite{Bittleston:2020hfv, Penna:2020uky,He:2021xoo,Cole:2023umd, Cole:2024sje}.

		At a higher conceptual level, we expect the boundary/pole structure of semi-holomophic Chern--Simons to be most naturally captured by the relative $L_\infty$-algebra perspective given in~\cite{Alfonsi:2024utl}. In particular, a relative $L_\infty$-algebra is necessary in the presence of space-times with boundary (and, more generally, corners). Essentially, the naive $L_\infty$-algebra associated with the theory in the absence of a boundary, is replaced by a pair of bulk and boundary $L_\infty$-algebras related by an $L_\infty$-morphism. This structure forms a mild generalisation of the Batalin--(Fradkin)--Vilkovisky construction~\cite{Cattaneo:2012qu} in the perturbative setting.\footnote{This ought to be replaced by a relative Batalin--Vilkovisky algebroid more generally.} The boundary $L_\infty$-algebra serves two purposes: (i) to ensure cyclicity whilst preserving the correct cohomology and (ii) to adjust the homotopy Maurer--Cartan action so that the trivial part of the S-matrix resides in the minimal model.\footnote{This is particularly important in the context of e.g.~Witten diagrams~\cite{Alfonsi:2024utl}.} The minimal model of the relative $L_\infty$-algebra is then a pair of quasi-isomorphic minimal models and one can therefore regard the physical degrees of freedom as residing in the boundary. Bulk configurations are encoded in the boundary through homotopy transfer.\footnote{If one then goes a step further and conjectures that the boundary theory is a dual conformal field theory, then we are in the realm of holography realised through homotopy transfer~\cite{Chiaffrino:2023wxk,Alfonsi:2024utl}.}

		In the context of semi-holomorphic Chern--Simons theory, this point of view suggests that we ought to regard $z_0\times\IC P^1$, where $z_0$ is the pole of the meromorphic one-form, as the boundary to which the four-dimensional semi-holomorphic Chern--Simons theory is transferred. To make proper sense of this, we must resolve $z_0 \times \IC P^1$ into a boundary proper (as opposed to a  topological boundary), $S^1\times \IC P^1$. Then, we may construct a relative $L_\infty$-algebra, where the choice of boundary conditions is given by a topological Batalin--Vilkovisky theory on the boundary as in~\cite{Kapustin:2009av,DiPietro:2019hqe}. The expectation is that upon applying homotopy transfer to the boundary and then taking a suitable $R\to 0$ limit, where $R$ is the $S^1$ radius, we shall recover the corresponding two-dimensional theory given by the choice of pole(s) and boundary conditions. This is suggestive of a novel realisation of dualities amongst the two-dimensional theories in terms of their four-dimensional parents and relative $L_\infty$-algebras. Prior to taking any limit we can regard the boundary theory on $S^1\times\IC P^1$ as a duality wall in the sense of~\cite{Kapustin:2006pk,Fuchs:2007fw,Gaiotto:2008ak, Kapustin:2009av} for dual four-dimensional Chern--Simons theories. The $R\to0$ ($R\to\infty$) limits then reduce to the dual two-dimensional theories.

		\section{Principal chiral model from semi-holomorphic Chern--Simons theory}\label{sec:PCMfromSHCS}

		\paragraph{Setting.}
		Let us first briefly review the setting following in parts~\cite{Costello:2017dso,Costello:2019tri,Delduc:2019whp,Benini:2020skc} (see also~\cite{Lacroix:2021iit} for a review). In particular, let $\Sigma$ be an oriented two-dimensional manifold. We consider $\Sigma\times\IC P^1$ coordinatised by $x^\mu$ with $\mu,\nu,\ldots=0,1$ on $\Sigma$ and $z$ a choice of global holomorphic coordinate on $\IC P^1$ together with the meromorphic one-form
		\begin{equation}\label{eq:Omega}
			\Omega\ \coloneqq\ \frac{1-z^2}{z^2}\,\rmd z~.
		\end{equation}
		Working with a compact semi-simple gauge group $\sfG$ with the associated Lie algebra $(\frg,[-,-])$ with metric $\inner{-}{-}$, the action of semi-holomorphic Chern--Simons theory is\footnote{The subscript `F' refers to F-structure in the sense of~\cite{Rawnsley} which is a generalisation of a Cauchy--Riemann structure to allow for real directions in the defining distribution. The associated Chern--Simons theory with $\Omega$ holomorphic was called partially holomorphic Chern--Simons theory in~\cite{Popov:2005uv}.}
		\begin{subequations}\label{eq:semiHoloCSTheory}
			\begin{equation}
				S^{\rm shCS}\ \coloneqq\ \frac{\rmi}{2\pi}\int_{\Sigma\times\IC P^1}\Omega\wedge\rmC\rmS_\rmF(A)~,
			\end{equation}
			where `i' is the imaginary unit and
			\begin{equation}
				\rmC\rmS_\rmF(A)\ \coloneqq\ \tfrac12\inner{A}{\bar\partial_\rmF A}+\tfrac1{3!}\inner{A}{[A,A]}
			\end{equation}
			with
			\begin{equation}
				\bar\partial_\rmF\ \coloneqq\ \underbrace{\rmd x^\mu\partial_\mu}_{\eqqcolon\,\rmd_\Sigma}+\underbrace{\rmd\bar z\partial_{\bar z}}_{\eqqcolon\,\bar\partial_{\IC P^1}}
				\eand
				A\ =\ \rmd x^\mu A_\mu+\rmd\bar zA_{\bar z}~.
			\end{equation}
			We assume $A_{\IC P^1}\coloneqq\rmd\bar zA_{\bar z}$ to be gauge-trivial and the boundary conditions
			\begin{equation}\label{eq:semiHoloBC}
				A_\Sigma|_{z\in\{0,\infty\}}\ =\ 0
				\ewith
				A_\Sigma\ \coloneqq\ \rmd x^\mu A_\mu
			\end{equation}
		\end{subequations}
		at the poles of $\Omega$.

		Next, we consider the field redefinition
		\begin{subequations}
			\begin{equation}
				L\ \coloneqq\ \hat g^{-1}A\hat g+\hat g^{-1}\bar\partial_\rmF\hat g~,
			\end{equation}
			where $\hat g$ is a smooth $\sfG$-valued function on $\Sigma\times\IC P^1$ that depends on $z,\bar z$ only through the modulus $|z|$ such that
			\begin{equation}\label{eq:defOfg}
				\hat g(x,|z|)\ =\
				\begin{cases}
					g(x) &\eforall|z|\ \leq\ r_1\eand x\ \in\ \Sigma
					\\
					\unit &\eforall|z|\ \geq\ r_2\ >\ r_1\eand x\ \in\ \Sigma
				\end{cases}
			\end{equation}
		\end{subequations}
		for some positive real numbers $r_1$ and $r_2$ and $g:\Sigma\rightarrow\sfG$ is arbitrary, and for $|z|\in[r_1,r_2]$, the function $\hat g$ interpolates smoothly between $g$ and $\unit$. Upon decomposing $L$ as
		\begin{equation}
			L\ \coloneqq\ J+a
			\ewith
			J\ \coloneqq\ \rmd x^\mu J_\mu
			\eand
			a\ \coloneqq\ \rmd\bar za_{\bar z}~,
		\end{equation}
		using the identities
		\begin{equation}\label{eq:deltaFunctionIdentities}
			\partial_{\bar z}\frac1z\ =\ -2\pi\rmi\delta^{(2)}(z)
			\quad\Rightarrow\quad
			\bar\partial_{\IC P^1}\Omega\ =\ -2\pi\rmi\partial_z\delta^{(2)}(z)\rmd z\wedge\rmd\bar z~,
		\end{equation}
		the action~\eqref{eq:semiHoloCSTheory} becomes
		\begin{subequations}\label{eq:redefinedSemiHoloCS}
			\begin{equation}\label{eq:redefinedSemiHoloCSAction}
				S^{\rm shCS}\ =\ \frac{\rmi}{2\pi}\int_{\Sigma\times\IC P^1}\Omega\wedge\big\{\tfrac12\inner{J}{\bar\partial_{\IC P^1}J}+\inner{a}{F_\Sigma(J)}\big\}-\frac12\int_\Sigma\inner{j}{\partial_z|_{z=0}J}
			\end{equation}
			where
			\begin{equation}
				F_\Sigma(J)\ \coloneqq\ \rmd_\Sigma J+\tfrac12[J,J]
				\eand
				j\ \coloneqq\ g^{-1}\rmd_\Sigma g~.
			\end{equation}
			Furthermore, the boundary conditions in~\eqref{eq:semiHoloBC} become
			\begin{equation}\label{eq:semiHoloBC2}
				J|_{z=0}\ =\ j
				\eand
				J|_{z=\infty}=0~.
			\end{equation}
		\end{subequations}
		Because of the boundary condition $J|_{z=0}=j$, we may rewrite the action~\eqref{eq:redefinedSemiHoloCSAction} as
		\begin{subequations}
			\begin{equation}\label{eq:redefinedSemiHoloCSAction2}
				S^{\rm shCS}\ =\ \frac{\rmi}{2\pi}\int_{\Sigma\times\IC P^1}\Omega\wedge\big\{\tfrac12\inner{J}{\bar\partial_{\IC P^1}J}+\inner{a}{F_\Sigma(J)}\big\}-\frac12\int_\Sigma\inner{J}{\partial_zJ}|_{z=0}
			\end{equation}
			which is supplemented by the boundary condition
			\begin{equation}\label{eq:refinedSemiHoloBC2}
				J|_{z=\infty}\ =\ 0~.
			\end{equation}
		\end{subequations}
		In doing so, the boundary condition of $J$ at $z=0$ will arise dynamically from the zero-curvature condition on $\Sigma$, as we shall see momentarily.

		\paragraph{On-shell configurations.}
		Thus far, we have merely summarised the standard argument. At this point, the next step in that argument is to solve the equations of motion
		\begin{equation}\label{eq:semiHoloEOM}
			\Omega\wedge\big\{\bar\partial_{\IC P^1}J+\rmd_\Sigma a+[J,a]\big\}\ =\ 0
			\eand
			\Omega\wedge F_\Sigma(J)\ =\ 0
		\end{equation}
		for~\eqref{eq:redefinedSemiHoloCSAction2}. Indeed, using~\eqref{eq:deltaFunctionIdentities}, it is now not too difficult to see that the most general solution to~\eqref{eq:semiHoloEOM} in the gauge in which $a=0$ and subject to the boundary condition~\eqref{eq:refinedSemiHoloBC2} is given by
		\begin{subequations}\label{eq:solSemiHoloCS}
			\begin{equation}
				J\ =\ \frac{1}{1-z^2}\,j+\frac{z}{1-z^2}\,\tilde j
				\eand
				a\ =\ 0
			\end{equation}
			with $\frg$-valued one-forms $j$ and $\tilde j$ on $\Sigma$ subject to
			\begin{equation}\label{eq:conditionsAlphaBeta}
				\rmd_\Sigma j+\tfrac12[j,j]\ =\ 0~,
				\quad
				\rmd_\Sigma\tilde j\ =\ 0~,
				\quad
				[j,j]+[\tilde j,\tilde j]\ =\ 0~,
				\quad
				[j,\tilde j]\ =\ 0~.
			\end{equation}
		\end{subequations}
		Note that $J|_{z=0}=j$ and from the first equation in~\eqref{eq:conditionsAlphaBeta} it follows that $j=g^{-1}\rmd_\Sigma g$ for some $g:\Sigma\rightarrow\sfG$ if the first de Rham cohomology group of $\Sigma$ is assumed to be zero. Hence, as stated before, the boundary condition $J|_{z=0}=j$ arises dynamically. If now, in addition, we restrict to on-shell configurations such that
		\begin{equation}\label{eq:duality}
			-\tfrac1z J\big(\tfrac1z\big)\ =\ {\star J(z)}~,
		\end{equation}
		where `$\star$' is the Hodge operator for an arbitrarily chosen Lorentzian metric $h$ on $\Sigma$, then $\tilde j={\star j}$ and~\eqref{eq:solSemiHoloCS} reduces further to
		\begin{equation}\label{eq:reducedSolSemiHoloCS}
			J\ =\ \frac{1}{1-z^2}\,j+\frac{z}{1-z^2}\,{\star j}~,
			\quad
			\rmd_\Sigma j+\tfrac12[j,j]\ =\ 0~,
			\quad
			\rmd_\Sigma{\star j}\ =\ 0~,
			\eand
			a\ =\ 0~.
		\end{equation}
		Upon evaluating~\eqref{eq:redefinedSemiHoloCSAction2} on this solution but without imposing $\rmd_\Sigma{\star j}=0$, we find
		\begin{equation}\label{eq:PCMAction}
			S^{\rm PCM}\ \coloneqq\ -\frac12\int_\Sigma\inner{j}{\star j}
			\ewith
			j\ =\ g^{-1}\rmd_\Sigma g~.
		\end{equation}
		This is the principal chiral model action on $(\Sigma,h)$ which has $\rmd_\Sigma{\star j}=0$ as its equation of motion. In this context, the one-form $J$ constitutes the Lax connection.

		It should be noted that the condition~\eqref{eq:duality} is an incarnation of the (pseudo-)duality of the principal chiral model discussed in~\cite{Zakharov:1973pp,Nappi:1979ig,Fridling:1983ha,Fradkin:1984ai,Curtright:1994be,Sarisaman:2007dm,Ricci:2007eq,Beisert:2008iq}. Alternatively, this condition can also be obtained by imposing certain boundary condition from the outset, see e.g.~\cite{Lacroix:2021iit} in light-cone coordinates for our choice of of volume form~\eqref{eq:Omega} and below in~\eqref{eq:shCSBDYConditions} in general coordinates.

		\paragraph{Remark.}
		Before moving on, it is amusing to note that when $\sfG=\sfS\sfU(2)$, the condition~\eqref{eq:duality} is no restriction on the possible solutions since in this case, there exists a Lorentzian conformal structure $[h]$ on $\Sigma$ such that~\eqref{eq:duality} holds automatically.

		To see this, we set $j_\pm\coloneqq\frac12(j\pm\tilde j)$ so that the two algebraic conditions in~\eqref{eq:conditionsAlphaBeta} are equivalent to
		\begin{equation}
			[j_\pm,j_\pm]\ =\ 0~.
		\end{equation}
		For $\frg=\frs\fru(2)$, this implies that the two components of $j_\pm$ must be proportional. Indeed, we may write $j_\pm=j^a_{\pm\mu}\sfe_a\otimes\rmd x^\mu$ with $\sfe_a$ for $a,b,\ldots=1,2,3$ a basis for $\frs\fru(2)$. Then, $[j_\pm,j_\pm]=0$ reads explicitly as $j_{0\pm}^aj_{\pm1}^b\eps_{abc}=0$ with $\eps_{abc}$ the Levi-Civita symbol. Consequently, $j_{\pm0}^a$ and $j_{\pm1}^a$ must be proportional. As we shall explain next, this implies that there is a Lorentzian conformal structure\footnote{On a two-dimensional Lorentzian manifold, we have $\star^2=1$ on one-forms and the (anti-)self-duality equation is conformally invariant.} $[h]$ on $\Sigma$ such that ${\star j_\pm}=\pm j_\pm$.

		The metric $h$ now arises as follows. Since the two components of $j_\pm$ are proportional, without loss of generality, we may write $j_\pm=a_\pm\otimes(\rmd x^0+b_\pm\rmd x^1)$ with $a_\pm\in\frs\fru(2)\otimes\Omega^0(\Sigma)$ and $b_\pm\in\Omega^0(\Sigma)$ with $b_+\neq b_-$.\footnote{For $b_+=b_-$, the differential conditions in~\eqref{eq:conditionsAlphaBeta}, which are the remaining conditions to be satisfied, are equivalent to $\rmd_\Sigma j_\pm=0$. In turn, this says that $\rmd_\Sigma j=0$ and $\rmd_\Sigma\tilde j=0$ and so, we obtain the trivial solution $(J,a)=(\rmd_\Sigma C,0)$ with $C\coloneqq\frac{1}{1-z^2}\,c+\frac{z}{1-z^2}\,\tilde c$ and $c,\tilde c\in\frs\fru(2)\otimes\Omega^0(\Sigma)$ if the first de Rham cohomology group of $\Sigma$ is assumed to be zero.} We now may change coordinates $x=(x^0,x^1)\mapsto y=(y^0,y^1)$ such that
		\begin{equation}
			\rmd x^0+b_\pm(x)\rmd x^1\ =\ c_\pm(y)(\rmd y^0\pm\rmd y^1)~,
		\end{equation}
		where $c_\pm\in\Omega^0(\Sigma)$ is non-vanishing. This system of partial differential equations always has a (local) solution for $b_+\neq b_-$. Therefore, in these coordinates, using $h=\diag(-1,1)$ as a representative of $[h]$, we have ${\star j_\pm}=\pm j_\pm$.

		Finally, the condition ${\star j_\pm}=\pm j_\pm$ implies that $\tilde j={\star j}$ and so, the differential conditions in~\eqref{eq:conditionsAlphaBeta}, which are the remaining conditions to be satisfied, become
		\begin{equation}
			\rmd_\Sigma j+\tfrac12[j,j]\ =\ 0
			\eand
			\rmd_\Sigma{\star j}\ =\ 0~.
		\end{equation}
		Altogether, we arrive at the solution~\eqref{eq:reducedSolSemiHoloCS} without having to invoke~\eqref{eq:duality} separately, that is,~\eqref{eq:duality} holds automatically.

		\section{\texorpdfstring{$L_\infty$}{L Infinity}-structure}\label{sec:LinftyStructure}

		\subsection{Semi-holomorphic Chern--Simons theory}\label{ssec:LinftyStructureCS}

		Let us now establish the cyclic $L_\infty$-structure underlying the semi-holomorphic Chern--Simons theory governed by~\eqref{eq:semiHoloCSTheory}. We shall directly incorporate all the necessary boundary conditions that eventually will allow us to construct a quasi-isomorphism between the cyclic $L_\infty$-algebras governing semi-holomorphic Chern--Simons theory and the principal chiral model. As before, we work with an oriented two-dimensional Lorentzian manifold $\Sigma$ with metric $h$.\footnote{Note that we do not assume the vanishing of the first de Rham cohomology group of $\Sigma$ in the following.}

		\paragraph{(Anti-)field spaces.}
		Recall that the gauge potential $A$ in~\eqref{eq:semiHoloCSTheory} decomposes as $A=A_\Sigma+A_{\IC P^1}$ with $A_\Sigma\coloneqq\rmd x^\mu A_\mu$ and $A_{\IC P^1}\coloneqq\rmd\bar zA_{\bar z}$, respectively. This is accompanied by the corresponding anti-field $A^+=A_\Sigma^++A_{\IC P^1}^+$ with $A_\Sigma^+\coloneqq\rmd\bar z\wedge\rmd x^\mu A^+_{\bar z\mu}$ and $A_{\IC P^1}^+\coloneqq\tfrac12\rmd x^\mu\wedge\rmd x^\nu A^+_{\mu\nu}$. We can decompose $A_\Sigma$ and $A_\Sigma^+$ into the self-dual and anti-self-dual parts along $\Sigma$ using the Hodge operator induced by $h$; we shall use the notation $\Omega^1_\pm(\Sigma)$ to indicate the self-dual and anti-self-dual one-forms on $\Sigma$. In addition, we have the ghost $c$ and its anti-field $c^+\coloneqq\tfrac12\rmd\bar z\wedge\rmd x^\mu\wedge\rmd x^\nu c^+_{\bar z\mu\nu}$. We now impose the boundary/pole conditions
		\begin{equation}\label{eq:shCSBDYConditions}
			\begin{gathered}
				c\ :\
				\begin{cases}
					c|_{z\in\{0,\infty\}}\ =\ 0
					\\
					c\ \in\ \frg\otimes\Omega^0(\Sigma)\otimes\Omega^{0,0}(\IC P^1)
				\end{cases},
				\\
				A_{\Sigma,\,\pm}\ :\
				\begin{cases}
					A_{\Sigma,\,\pm}|_{z\in\{0,\infty\}}\ =\ 0
					\\
					A_{\Sigma,\,\pm}|_{\Sigma\times\IC P^1\setminus U_\pm}\ \in\ \frg\otimes\Omega^1_\pm(\Sigma)\otimes\Omega^{0,0}(\IC P^1\setminus U_\pm)
					\\
					A_{\Sigma,\,\pm}\big|_{\Sigma\times U_\pm}\ =\ \frac{1}{1\mp z}\alpha_\pm\ewith\alpha_\pm\ \in\ \frg\otimes\Omega^1_\pm(\Sigma)\otimes\Omega^{0,0}(U_\pm)
				\end{cases},
				\\
				A_{\IC P^1}\ :\
				\begin{cases}
					A_{\IC P^1}|_{z\in\{0,\infty\}}\ =\ 0
					\\
					A_{\IC P^1}\ \in\ \frg\otimes\Omega^0(\Sigma)\otimes\Omega^{0,1}(\IC P^1)
				\end{cases},
				\\
				A^+_{\Sigma,\,\pm}\ :\
				\begin{cases}
					A^+_{\Sigma,\,\pm}|_{z\in\{0,\infty\}}\ =\ 0
					\\
					A^+_{\Sigma,\,\pm}|_{\Sigma\times\IC P^1\setminus U_\pm}\ \in\ \frg\otimes\Omega^1_\pm(\Sigma)\otimes\Omega^{0,1}(\IC P^1\setminus U_\pm)
					\\
					A^+_{\Sigma,\,\pm}\big|_{\Sigma\times U_\pm}\ =\ \frac{1}{1\mp z}\alpha_\pm\ewith\alpha_\pm\ \in\ \frg\otimes\Omega^1_\pm(\Sigma)\otimes\Omega^{0,1}(U_\pm)
				\end{cases}~,
				\\
				A^+_{\IC P^1,\,\pm}\ :\
				\begin{cases}
					A^+_{\IC P^1,\,\pm}|_{z\in\{0,\infty\}}\ =\ 0
					\\
					A^+_{\IC P^1,\,\pm}|_{\Sigma\times\IC P^1\setminus U_\pm}\ \in\ \frg\otimes\Omega^2_\pm(\Sigma)\otimes\Omega^{0,0}(\IC P^1\setminus U_\pm)
					\\
					A^+_{\IC P^1,\,\pm}\big|_{\Sigma\times U_\pm}\ =\ \frac{1}{1\mp z}\alpha_\pm\ewith\alpha_\pm\ \in\ \frg\otimes\Omega^2_\pm(\Sigma)\otimes\Omega^{0,0}(U_\pm)
				\end{cases},
				\\
				c^+\, :\,
				\begin{cases}
					c^+|_{z\in\{0,\infty\}}\ =\ 0
					\\
					c^+|_{\Sigma\times\IC P^1\setminus U_\pm}\ \in\ \frg\otimes\Omega^2_\pm(\Sigma)\otimes\Omega^{0,1}(\IC P^1\setminus U_\pm)
					\\
					c^+\big|_{\Sigma\times U_\pm}\ =\ \frac{1}{1\mp z}\alpha_\pm\ewith\alpha_\pm\ \in\ \frg\otimes\Omega^2_\pm(\Sigma)\otimes\Omega^{0,1}(U_\pm)
				\end{cases},
			\end{gathered}
		\end{equation}
		where $U_\pm\subseteq\IC P^1$ are open neighbourhoods of $z=\pm1$.\footnote{It should be noted that the boundary/pole conditions~\eqref{eq:shCSBDYConditions} are a covariantisation of the conditions given e.g.~in~\cite{Lacroix:2021iit}. However, they differ from those used e.g.~in~\cite[Example 3.10]{Benini:2026pwa}. Nevertheless, one can show that both types of boundary/pole conditions lead to quasi-isomorphic complexes and so, the corresponding theories are also quasi-isomorphic.} We shall denote the corresponding (anti-)field spaces by $\frL^{\rm shCS}_0\ni c$ and $\frL^{\rm shCS}_3\ni c^+$, $\frL^{\rm shCS}_1\coloneqq\frL^{\rm shCS}_{1,\,\Sigma}\oplus\frL^{\rm shCS}_{1,\,\IC P^1}\ni(A_\Sigma,A_{\IC P^1})$, and $\frL^{\rm shCS}_2\coloneqq\frL^{\rm shCS}_{2,\,\Sigma}\oplus\frL^{\rm shCS}_{2,\,\IC P^1}\ni(A^+_\Sigma,A^+_{\IC P^1})$, respectively. It should be noted that these boundary/pole condition ultimately ensure the on-shell (pseudo-)duality~\eqref{eq:duality}.

		\paragraph{$L_\infty$-algebra $\frL^{\rm shCS}$.}
		From~\eqref{eq:shCSBDYConditions} it is evident that all our fields and anti-fields are assumed to have a finite number of poles. In general, let $(a_1,\ldots,a_k)$ be the set of all poles of orders $(n_1,\ldots,n_k)$ of a $(0,p)$-form $\alpha$ on $\IC P^1$. We define a regularised anti-holomorphic derivative
		\begin{equation}\label{eq:regularisedAntiHoloDer}
			\bar\partial'_{\IC P^1}\alpha\big|_{z=z_0}\ \coloneqq\
			\begin{cases}
				\bar\partial_{\IC P^1}\alpha\big|_{z=z_0} & \efor z_0\not\in(a_1,\ldots,a_k)
				\\
				\frac{1}{(z-a_k)^{n_k}}\bar\partial_{\IC P^1}[(z-a_k)^{n_k}\alpha]\bigg|_{z_0=a_k} & \eelse
			\end{cases}
		\end{equation}
		on $\IC P^1$, and it is not too difficult to see that this is a differential derivation.

		We now set $\frL^{\rm shCS}\coloneqq\bigoplus_{k=0}^3\frL^{\rm shCS}_k$ with
		\begin{subequations}\label{eq:shCSLInftyAlgebra}
			\begin{equation}\label{eq:shCSComplex}
				\sfCh(\frL^{\rm shCS})\ \coloneqq\
				\left(
					\begin{tikzcd}
						\overbrace{\frL^{\rm shCS}_0}^{\ni\,c}\rar["\mu_1"] & \overbrace{\frL^{\rm shCS}_{1,\,\Sigma}\oplus\frL^{\rm shCS}_{1,\,\IC P^1}}^{\ni\,(A_\Sigma,A_{\IC P^1})}\rar["\mu_1"] & \overbrace{\frL^{\rm shCS}_{2,\,\Sigma}\oplus\frL^{\rm shCS}_{2,\,\IC P^1}}^{\ni\,(A_\Sigma^+,A_{\IC P^1}^+)}\rar["\mu_1"] & \overbrace{\frL^{\rm shCS}_3}^{\ni\,c^+}
					\end{tikzcd}
				\right)
			\end{equation}
			and the differential
			\begin{equation}
				\begin{gathered}
					\mu^{\rm shCS}_1(c)\ \coloneqq\
					\begin{pmatrix}
						\rmd_\Sigma c\\ \bar\partial'_{\IC P^1}c
					\end{pmatrix},
					\quad
					\mu^{\rm shCS}_1
					\begin{pmatrix}
						A_\Sigma \\ A_{\IC P^1}
					\end{pmatrix}
					\ \coloneqq\
					\begin{pmatrix}
						\bar\partial'_{\IC P^1}A_\Sigma+\rmd_\Sigma A_{\IC P^1} \\ \rmd_\Sigma A_\Sigma
					\end{pmatrix},
					\\
					\mu^{\rm shCS}_1
					\begin{pmatrix}
						A^+_\Sigma \\ A^+_{\IC P^1}
					\end{pmatrix}
					\ \coloneqq\ \bar\partial'_{\IC P^1}A^+_{\IC P^1}+\rmd_\Sigma A^+_\Sigma
				\end{gathered}
			\end{equation}
			as well as the higher products
			\begin{equation}
				\begin{gathered}
					\mu^{\rm shCS}_2(c,c')\ \coloneqq\ [c,c']~,
					\quad
					\mu^{\rm shCS}_2(c,c^+)\ \coloneqq\ [c,c^+]~,
					\\
					\mu^{\rm shCS}_2
					\left(
						c,
						\begin{pmatrix}
							A_\Sigma\\ A_{\IC P^1}
						\end{pmatrix}
					\right)
					\ \coloneqq\
					\begin{pmatrix}
						[c,A_\Sigma]\\ [c,A_{\IC P^1}]
					\end{pmatrix},
					\quad
					\mu^{\rm shCS}_2
					\left(
						c,
						\begin{pmatrix}
							A^+_\Sigma\\ A^+_{\IC P^1}
						\end{pmatrix}
					\right)
					\ \coloneqq\
					\begin{pmatrix}
						[c,A^+_\Sigma] \\ [c,A^+_{\IC P^1}]
					\end{pmatrix},
					\\
					\mu^{\rm shCS}_2
					\left(
						\begin{pmatrix}
							A_\Sigma \\ A_{\IC P^1}
						\end{pmatrix},
						\begin{pmatrix}
							A'_\Sigma\\ A'_{\IC P^1}
						\end{pmatrix}
					\right)
					\ \coloneqq\
					\begin{pmatrix}
						[A_{\IC P^1},A'_\Sigma]+[A_\Sigma,A'_{\IC P^1}] \\ [A_\Sigma,A'_\Sigma]
					\end{pmatrix},
					\\
					\mu^{\rm shCS}_2
					\left(
						\begin{pmatrix}
							A_\Sigma \\ A_{\IC P^1}
						\end{pmatrix},
						\begin{pmatrix}
							A^+_\Sigma \\ A^+_{\IC P^1}
						\end{pmatrix}
					\right)
					\ \coloneqq\ [A_\Sigma,A^+_\Sigma]+[A_{\IC P^1},A^+_{\IC P^1}]~.
				\end{gathered}
			\end{equation}
		\end{subequations}
		It is not too difficult to check that $(\frL^{\rm shCS},\mu^{\rm shCS}_i)$ forms an $L_\infty$-algebra. Note that since the wedge product of two self-dual or anti-self-dual forms vanishes, we never produce higher order poles in these products and, consequently, the boundary/pole conditions~\eqref{eq:shCSBDYConditions} are compatible with these products.

		\paragraph{Cyclic structure.}
		In addition, we introduce the non-degenerate graded symmetric bilinear form
		\begin{equation}\label{eq:shCSCyclicStructure}
			\inner{\phi}{\phi^+}^{\rm shCS}\ \coloneqq\ \frac{\rmi}{2\pi}\int_{\Sigma\times\IC P^1}\Omega\wedge\inner{\phi}{\phi^+}
		\end{equation}
		for all $\phi\in\{c,A_\Sigma,A_{\IC P^1}\}$ and $\phi^+\in\{A^+_\Sigma,A^+_{\IC P^1},c^+\}$.\footnote{Alternatively, one could include $\Omega$ in the anti-fields (with suitable pole, zeros and factorisation conditions imposed), so that the cyclic structure is just the usual integral of top-forms induced by the wedge product.} This makes $(\frL^{\rm shCS},\mu^{\rm shCS}_i)$ into a cyclic $L_\infty$-algebra. Note that the boundary/pole structure~\eqref{eq:shCSBDYConditions}, especially the splitting of the poles of the self-dual and anti-self-dual parts, is essential for this bilinear form~\eqref{eq:shCSCyclicStructure} to be well-defined, as two (anti)-self-dual one-forms along $\Sigma$ cannot appear together in the bilinear form which avoids higher order poles. The non-degeneracy is due to the fact that if $\Omega\wedge\phi=0$ then $\phi=0$, where $\phi$ represents any element in the $L_\infty$-algebra. This is due to the smoothness properties and the fact that $\Omega$ has isolated zeros. Finally, since $\Omega$ is zero at $z=\pm 1$, the regularised version $\bar\partial'_{\IC P^1}$ defined in~\eqref{eq:regularisedAntiHoloDer} and the bare version $\bar\partial_{\IC P^1}$ of the anti-holomorphic exterior derivative behave the same when appearing in the bilinear form~\eqref{eq:shCSCyclicStructure}.

		\paragraph{Batalin--Vilkovisky action.}
		Using~\eqref{eq:shCSLInftyAlgebra} and~\eqref{eq:shCSCyclicStructure}, the homotopy Maurer--Cartan action for this $L_\infty$-algebra is the Batalin--Vilkovisky action
		\begin{equation}\label{eq:shCSBVAction}
			\begin{aligned}
				S^{\rm shCS,\,BV}\ &\coloneqq\ \frac{\rmi}{2\pi}\int_{\Sigma\times\IC P^1}\Omega\wedge\big\{\tfrac12\inner{A}{\bar\partial'_\rmF A}+\tfrac1{3!}\inner{A}{[A,A]}
				\\
				&\kern4cm-\inner{A^+}{\bar\partial'_\rmF c+[A,c]}+\tfrac12\inner{c^+}{[c,c]}\big\}
			\end{aligned}
		\end{equation}
		for semi-holomorphic Chern--Simons theory where $\bar\partial'_\rmF\coloneqq\rmd_\Sigma+\bar\partial'_{\IC P^1}$ with $\bar\partial'_{\IC P^1}$ given in~\eqref{eq:regularisedAntiHoloDer}.

		Note that because of the non-degeneracy of~\eqref{eq:shCSCyclicStructure}, the equations of motion for $A$ following from~\eqref{eq:shCSBVAction} are
		\begin{equation}
			\bar\partial'_{\IC P^1}A_\Sigma+\rmd_\Sigma A_{\IC P^1}+[A_\Sigma,A_{\IC P^1}]\ =\ 0
			\eand
			\rmd_\Sigma A_\Sigma+\tfrac12[A_\Sigma,A_\Sigma]\ =\ 0
		\end{equation}
		under the assumption of vanishing anti-fields; note that $\Omega$ does not appear here. This is contrary to~\eqref{eq:semiHoloEOM} and due to our choice of the boundary/pole structure~\eqref{eq:shCSBDYConditions} and the use of the regularised differential~\eqref{eq:regularisedAntiHoloDer}.

		\subsection{Principal chiral model}\label{sec:PCM}

		Let us now recap the $L_\infty$-structure underlying the principal chiral model on a Lorentzian manifold $(\Sigma,h)$ as discussed in~\cite{JalaliFarahani:2023sfq}.

		\paragraph{Action revisited.}
		Recall the action~\eqref{eq:PCMAction}. We may parametrise\footnote{The parametrisation $g=\rme^\phi$ is a change of variables that is valid locally, in a neighbourhood of the identity of $\sfG$, where the exponential map is a diffeomorphism. It is therefore a local (perturbative) parametrisation about the vacuum $g=\unit$, which is the setting in which the $L_\infty$-algebraic description over a linear space of fields applies. Global aspects of the principal chiral model, which stem from the non-linear nature of its field space, are not captured in the present approach.} $g:\Sigma\rightarrow\sfG$ as $g=\rme^\phi$ for $\phi:\Sigma\rightarrow\frg$. Then,
		\begin{equation}\label{eq:PCMCurrent}
			j\ =\ \sum_{n=0}^\infty\frac{(-1)^n}{(n+1)!}\,\ad_\phi^n(\rmd_\Sigma\phi)
		\end{equation}
		with $\ad_\phi(-)\coloneqq[\phi,-]$, and~\eqref{eq:PCMAction} becomes
		\begin{equation}\label{eq:PCMActionPhi}
			S^{\rm PCM}\ =\ -\sum_{n=0}^\infty\frac{1}{(2n+2)!}\int\inner{\rmd_\Sigma\phi}{\star\ad^{2n}_\phi(\rmd_\Sigma\phi)}~.
		\end{equation}
		In deriving this, we have noted that only even powers of $\ad_\phi$ contribute after inserting~\eqref{eq:PCMCurrent} and then made use of $\sum_{m=0}^{2n}\frac{(-1)^m}{(m+1)!(2n-m+1)!}=\frac{2}{(2n+2)!}$. We may rewrite~\eqref{eq:PCMActionPhi} further as
		\begin{equation}
			\begin{aligned}\label{eq:PCMActionPhiRewritten}
				S^{\rm PCM}\ &=\ -\sum_{n=0}^\infty\frac{1}{(2n+2)!}\frac{1}{2n+2}
				\\
				&\kern1cm\times\int\left\langle\phi,\star\left\{\sum_{m=0}^{2n-1}(-1)^m{\star[\ad^m_\phi(\rmd_\Sigma\phi),\star\ad^{2n-1-m}_\phi(\rmd_\Sigma\phi)]}+2\rmd_\Sigma^\dagger\ad^{2n}_\phi(\rmd_\Sigma\phi)\right\}\right\rangle
			\end{aligned}
		\end{equation}
		upon integration by parts. Here, $\rmd_\Sigma^\dagger\coloneqq{\star\rmd_\Sigma\star}$ is the adjoint of $\rmd_\Sigma$ when acting on one-forms.

		\paragraph{$L_\infty$-algebra $\frL^{\rm PCM}$.}
		We have the field space $\frL_1^{\rm PCM}\ni\phi$ and the corresponding anti-field space $\frL_2^{\rm PCM}\ni\phi^+$ with $\phi,\phi^+$ smooth $\frg$-valued functions on $\Sigma$. We then set $\frL^{\rm PCM}\coloneqq\frL_1^{\rm PCM}\oplus\frL_2^{\rm PCM}$ and
		\begin{subequations}\label{eq:PCMLInftyAlgebra}
			\begin{equation}\label{eq:PCMComplex}
				\sfCh(\frL^{\rm PCM})\ \coloneqq\
				\left(
					\begin{tikzcd}
						\overbrace{\frL^{\rm PCM}_1}^{\ni\,\phi}\rar["\mu_1"] & \overbrace{\frL^{\rm PCM}_2}^{\ni\,\phi^+}
					\end{tikzcd}
				\right)
			\end{equation}
			and the differential
			\begin{equation}
				\mu^{\rm PCM}_1(\phi)\ \coloneqq\ -\rmd_\Sigma^\dagger\rmd_\Sigma\phi
			\end{equation}
			as well as the higher products, which are obtained by polarising
			\begin{equation}
				\begin{aligned}
					\mu^{\rm PCM}_{2n+1}(\phi,\ldots,\phi)\ &\coloneqq\ -\frac{1}{2n+2}\left\{2\rmd_\Sigma^\dagger\ad^{2n}_\phi(\rmd_\Sigma\phi)\phantom{\sum_{m=0}^{2n-1}}\right.
					\\
					&\kern3cm\left.+\sum_{m=0}^{2n-1}(-1)^m{\star[\ad^m_\phi(\rmd_\Sigma\phi),\star\ad^{2n-1-m}_\phi(\rmd_\Sigma\phi)]}\right\}.
				\end{aligned}
			\end{equation}
		\end{subequations}
		It is not too difficult to check that $(\frL^{\rm PCM},\mu^{\rm PCM}_i)$ forms an $L_\infty$-algebra.

		\paragraph{Cyclic structure.}
		In addition, we introduce the non-degenerate graded symmetric bilinear form
		\begin{equation}\label{eq:PCMCyclicStructure}
			\inner{\phi}{\phi^+}^{\rm PCM}\ \coloneqq\ \int_\Sigma\inner{\phi}{\star\phi^+}
		\end{equation}
		making $(\frL^{\rm PCM},\mu^{\rm PCM}_i)$ into a cyclic $L_\infty$-algebra.

		The corresponding homotopy Maurer--Cartan action is then the action~\eqref{eq:PCMActionPhiRewritten} or~\eqref{eq:PCMActionPhi}. Note that the reason for rewriting~\eqref{eq:PCMActionPhi} as~\eqref{eq:PCMActionPhiRewritten} was to ensure that the polarised higher products in~\eqref{eq:PCMLInftyAlgebra} are indeed cyclic for~\eqref{eq:PCMCyclicStructure}; this would not be the case when directly reading off~\eqref{eq:PCMActionPhi} the higher products.

		\section{Quasi-isomorphism}\label{sec:quasi-iso}

		We recall from \cref{sec:PCMfromSHCS} that the principal chiral model arises from the semi-holomorphic Chern--Simons theory through a combination of partial elimination of the equations of motion by algebraic means (i.e.~no change in cohomology) and gauge fixing. Consequently, the two theories are expected to be quasi-isomorphic on general grounds. From the perspective of the Batalin--Vilkovisky formalism, this reduction corresponds to eliminating trivial pairs, or generalised auxiliary fields in the terminology of~\cite{Henneaux:1989ua,Barnich:2004cr,Krotov:2006th-arXiv,Krotov:2006th,Grigoriev:2010ic,Barnich:2010sw,Grigoriev:2012xg,Simao:2021xgw,Cattaneo:2023cnt}. Importantly, in this section, we provide an explicit cyclic (i.e.~off-shell) quasi-isomorphism that directly captures the Lax connection and thereby encodes the infinite hierarchy of conserved charges that underlies the integrability of the principal chiral model. This is the first step towards analysing this correspondence between these two theories at the quantum level via quasi-isomorphisms of quantum $L_\infty$-algebras and thus, analysing quantum integrability.

		For the ease of construction, we shall switch to the dual Chevalley--Eilenberg description of the $L_\infty$-algebras from \cref{sec:LinftyStructure}. See e.g.~\cite{Jurco:2018sby,Jurco:2019bvp} for details on switching between these two pictures. Note that the Chevalley--Eilenberg complex underlying the Chevalley--Eilenberg description of the $L_\infty$-algebras is also known as the Batalin--Vilkovisky complex in the physics literature~\cite{Henneaux:1992}.

		\subsection{Principal chiral model in the Chevalley--Eilenberg picture}

		Let us start with the principal chiral model as described in \cref{sec:PCM}.

		\paragraph{Chevalley–Eilenberg algebra $\sfA^{\rm PCM}$.}
		In~\eqref{eq:PCMLInftyAlgebra}, we have spelled out the $L_\infty$-algebra for the principal chiral model. In the dual language, the Chevalley--Eilenberg algebra is generated by the field $\check\phi$ and its anti-field $\check\phi^+$. The Chevalley--Eilenberg differential\footnote{This should be interpreted as the differential on the Batalin--Vilkovisky complex, or, equivalently, as the Koszul differential, since the principal chiral model has no gauge symmetry.} is then given by
		\begin{equation}
			\begin{aligned}
				\sfd_{\rm CE}^{\rm PCM}\check\phi\ &\coloneqq\ 0~,
				\\
				\sfd_{\rm CE}^{\rm PCM}\check\phi^+\ &\coloneqq\ \sum_{n=0}^\infty\frac{1}{(2n+2)!}\left\{2\rmd_\Sigma^\dagger\ad^{2n}_{\check\phi}(\rmd_\Sigma\check\phi)\phantom{\sum_{m=0}^{2n-1}}\right.
				\\
				&\kern4cm\left.+\sum_{m=0}^{2n-1}(-1)^m{\star[\ad^m_{\check\phi}(\rmd_\Sigma\check\phi),\star\ad^{2n-1-m}_{\check\phi}(\rmd_\Sigma\check\phi)]}\right\}.
			\end{aligned}
		\end{equation}
		We denote the corresponding Chevalley--Eilenberg algebra by $\sfA^{\rm PCM}$.

		To simplify our subsequent discussion, let us perform the sums explicitly. First of all, we note that $[\alpha,{\star\beta}]=-[\beta,{\star\alpha}]$ by the properties of the Hodge operator. Therefore, it is not too difficult to see that $\sum_{m=0}^{2n}(-1)^m{\star[\ad^m_{\check\phi}(\rmd_\Sigma\check\phi),\star\ad^{2n-m}_{\check\phi}(\rmd_\Sigma\check\phi)]}=0$ and so, we may add this sum to $\sfd_{\rm CE}^{\rm PCM}\check\phi^+$ to obtain
		\begin{equation}
			\begin{aligned}
				\sfd_{\rm CE}^{\rm PCM}\check\phi^+\ &=\ \sum_{n=0}^\infty\frac{2}{(2n+2)!}\rmd_\Sigma^\dagger\ad^{2n}_{\check\phi}(\rmd_\Sigma\check\phi)
				\\
				&\kern1cm+\sum_{n=0}^\infty\frac{1}{(n+3)!}\sum_{m=0}^n(-1)^m{\star[\ad^m_{\check\phi}(\rmd_\Sigma\check\phi),\star\ad^{n-m}_{\check\phi}(\rmd_\Sigma\check\phi)]}~.
			\end{aligned}
		\end{equation}
		Next, recall that $\int_0^1\rmd t\,(1-t)^kt^l=\frac{k!l!}{(k+l+1)!}$ and since $n+3=(m+1)+(n-m+1)+1$, we obtain
		\begin{equation}\label{eq:CEPCMRewritten}
			\begin{aligned}
				\sfd_{\rm CE}^{\rm PCM}\check\phi^+\ &=\ \sum_{n=0}^\infty\frac{2}{(2n+2)!}\rmd_\Sigma^\dagger\ad^{2n}_{\check\phi}(\rmd_\Sigma\check\phi)
				\\
				&\kern1cm+\int_0^1\rmd t\,(1-t)t\sum_{n=0}^\infty\sum_{m=0}^n\frac{(-1)^m{\star[\ad^m_{(1-t)\check\phi}(\rmd_\Sigma\check\phi),\star\ad^{n-m}_{t\check\phi}(\rmd_\Sigma\check\phi)]}}{(m+1)!(n-m+1)!}
				\\
				&=\ \left(\rmd_\Sigma^\dagger\left(\frac{\rme^{\ad_{\check\phi}}-\id}{\ad_{\check\phi}}\right)\left(\frac{\id-\rme^{-\ad_{\check\phi}}}{\ad_{\check\phi}}\right)\right)(\rmd_\Sigma\check\phi)
				\\
				&\kern1cm+\int_0^1\rmd t\,\star\!\left[\left(\frac{\id-\rme^{-(1-t)\ad_{\check\phi}}}{\ad_{\check\phi}}\right)(\rmd_\Sigma\check\phi), \left(\frac{\rme^{t\ad_{\check\phi}}-\id}{\ad_{\check\phi}}\right)({\star\rmd_\Sigma\check\phi})\right]
				\\
				&=\ \left(\rmd_\Sigma^\dagger\left(\frac{\rme^{\ad_{\check\phi}}-\id}{\ad_{\check\phi}}\right)\left(\frac{\id-\rme^{-\ad_{\check\phi}}}{\ad_{\check\phi}}\right)\right)(\rmd_\Sigma\check\phi)
				\\
				&\kern1cm-\int_0^1\rmd t\,\star\rme^{t\ad_{\check\phi}}\left[\left(\frac{\id-\rme^{-t\ad_{\check\phi}}}{\ad_{\check\phi}}\right)(\rmd_\Sigma\check\phi), \left(\frac{\id-\rme^{-\ad_{\check\phi}}}{\ad_{\check\phi}}\right)({\star\rmd_\Sigma\check\phi})\right],
			\end{aligned}
		\end{equation}
		where in the second step we have used the Cauchy product formula for infinite series and in the last step we have subtracted and added $\rme^{t\ad_{\check\phi}}$ in the first slot of the commutator and used $[\alpha,{\star\beta}]=-[\beta,{\star\alpha}]$ together with the fact that $\rme^{\ad_X}([\alpha,\beta])=[\rme^{\ad_X}(\alpha),\rme^{\ad_X}(\beta)])$.

		\paragraph{Chevalley–Eilenberg algebra $\sfA'^{\rm PCM}$.}
		Alternatively, we may describe the principal chiral model by the same graded algebra but with the differential
		\begin{subequations}\label{eq:CEPCMHat}
			\begin{equation}
				\sfd'^{\rm PCM}_{\rm CE}\check\phi'\ \coloneqq\ 0
				\eand
				\sfd'^{\rm PCM}_{\rm CE}\check\phi'^+\ \coloneqq\ \rmd_\Sigma^\dagger\check j'
			\end{equation}
			with
			\begin{equation}
				\check j'\ \coloneqq\ \rme^{-\check\phi'}\rmd_\Sigma\rme^{\check\phi'}\ =\ \sum_{n=0}^\infty\frac{(-1)^n}{(n+1)!}\ad_{\check\phi'}^n(\rmd_\Sigma\check\phi')\ =\ \left(\frac{\id-\rme^{-\ad_{\check\phi'}}}{\ad_{\check\phi'}}\right)(\rmd_\Sigma\check\phi')~.
			\end{equation}
		\end{subequations}
		We denote the corresponding Chevalley--Eilenberg algebra by $\sfA'^{\rm PCM}$.

		In particular, we shall now establish an isomorphism between the Chevalley--Eilenberg algebras $\sfA^{\rm PCM}$ and $\sfA'^{\rm PCM}$, or put differently, an isomorphism of the corresponding $L_\infty$-algebras.

		Consider
		\begin{equation}\label{eq:PCMPCMPrimeMorphism}
			\begin{aligned}
				\Phi\,:\,\sfA^{\rm PCM}\ &\rightarrow\ \sfA'^{\rm PCM}~,
				\\
				\check\phi\ &\mapsto\ \check\phi'~,
				\\
				\check\phi^+\ &\mapsto\ \left(\frac{\rme^{\ad_{\check\phi'}}-\id}{\ad_{\check\phi'}}\right)(\check\phi'^+)\ =\ \sum_{n=0}^\infty\frac{1}{(n+1)!}\ad_{\check\phi'}^n(\check\phi'^+)~.
			\end{aligned}
		\end{equation}
		Evidently, this is an isomorphism of graded algebras with the inverse given by
		\begin{equation}
			\begin{aligned}
				\Phi^{-1}\,:\,\sfA'^{\rm PCM}\ &\rightarrow\ \sfA^{\rm PCM}~,
				\\
				\check\phi'\ &\mapsto\ \check\phi~,
				\\
				\check\phi'^+\ &\mapsto\ \left(\frac{\ad_{\check\phi}}{\rme^{\ad_{\check\phi}}-\id}\right)(\check\phi^+)\ =\ \sum_{n=0}^\infty\frac{B_n}{n!}\ad_{\check\phi}^n(\check\phi^+)
			\end{aligned}
		\end{equation}
		with $B_n$ the Bernoulli numbers with normalisation $B_1=-\frac12$.

		Next, we show that
		\begin{equation}\label{eq:CEPhiPCMCom}
			\sfd'^{\rm PCM}_{\rm CE}\circ\Phi\ =\ \Phi\circ\sfd^{\rm PCM}_{\rm CE}~,
		\end{equation}
		that is, $\Phi$ is, in fact, an isomorphism of differential graded algebras. To this end, the only non-trivial part we need to verify is the action of this equation on the anti-field $\check\phi^+$. In particular, we obtain
		\begin{align}
			&(\sfd'^{\rm PCM}_{\rm CE}\circ\Phi)(\check\phi^+)\notag
			\\
			&=\ \left(\left(\frac{\rme^{\ad_{\check\phi'}}-\id}{\ad_{\check\phi'}}\right)\rmd_\Sigma^\dagger\left(\frac{\id-\rme^{-\ad_{\check\phi'}}}{\ad_{\check\phi'}}\right)\right)(\rmd_\Sigma\check\phi')\notag
			\\
			&=\ \left(\rmd_\Sigma^\dagger\left(\frac{\rme^{\ad_{\check\phi'}}-\id}{\ad_{\check\phi'}}\right)\left(\frac{\id-\rme^{-\ad_{\check\phi'}}}{\ad_{\check\phi'}}\right)-\left[\rmd_\Sigma^\dagger,\frac{\rme^{\ad_{\check\phi'}}-\id}{\ad_{\check\phi'}}\right]\left(\frac{\id-\rme^{-\ad_{\check\phi'}}}{\ad_{\check\phi'}}\right)\right)(\rmd_\Sigma\check\phi')\notag
			\\
			&=\ \left(\rmd_\Sigma^\dagger\left(\frac{\rme^{\ad_{\check\phi'}}-\id}{\ad_{\check\phi'}}\right)\left(\frac{\id-\rme^{-\ad_{\check\phi'}}}{\ad_{\check\phi'}}\right)\right)(\rmd_\Sigma\check\phi')\notag
			\\
			&\kern1cm-\int_0^1\rmd t\,\star\left(\left[\rmd_\Sigma,\rme^{t\ad_{\check\phi'}}\right]\left(\frac{\id-\rme^{-\ad_{\check\phi'}}}{\ad_{\check\phi'}}\right)({\star\rmd_\Sigma\check\phi'})\right)\notag
			\\
			&=\ \left(\rmd_\Sigma^\dagger\left(\frac{\rme^{\ad_{\check\phi'}}-\id}{\ad_{\check\phi'}}\right)\left(\frac{\id-\rme^{-\ad_{\check\phi'}}}{\ad_{\check\phi'}}\right)\right)(\rmd_\Sigma\check\phi')\notag
			\\
			&\kern1cm-\int_0^1\rmd t\int_0^1\rmd s\,\star\left(\rme^{(1-s)t\ad_{\check\phi'}}t\ad_{\rmd_\Sigma\check\phi'}\rme^{st\ad_{\check\phi'}}\left(\frac{\id-\rme^{-\ad_{\check\phi'}}}{\ad_{\check\phi'}}\right)({\star\rmd_\Sigma\check\phi'})\right)\notag
			\\
			&=\ \left(\rmd_\Sigma^\dagger\left(\frac{\rme^{\ad_{\check\phi'}}-\id}{\ad_{\check\phi'}}\right)\left(\frac{\id-\rme^{-\ad_{\check\phi'}}}{\ad_{\check\phi'}}\right)\right)(\rmd_\Sigma\check\phi')
			\\
			&\kern1cm-\int_0^1\rmd t\,\rme^{t\ad_{\check\phi'}}\int_0^t\rmd s\,\star\left(\rme^{-s\ad_{\check\phi'}}\ad_{\rmd_\Sigma\check\phi'}\rme^{s\ad_{\check\phi'}}\left(\frac{\id-\rme^{-\ad_{\check\phi'}}}{\ad_{\check\phi'}}\right)({\star\rmd_\Sigma\check\phi'})\right)\notag
			\\
			&=\ \left(\rmd_\Sigma^\dagger\left(\frac{\rme^{\ad_{\check\phi'}}-\id}{\ad_{\check\phi'}}\right)\left(\frac{\id-\rme^{-\ad_{\check\phi'}}}{\ad_{\check\phi'}}\right)\right)(\rmd_\Sigma\check\phi')\notag
			\\
			&\kern1cm-\int_0^1\rmd t\,\rme^{t\ad_{\check\phi'}}\int_0^t\rmd s\,\star\left[\rme^{-s\ad_{\check\phi'}}(\rmd_\Sigma\check\phi'),\left(\frac{\id-\rme^{-\ad_{\check\phi'}}}{\ad_{\check\phi'}}\right)({\star\rmd_\Sigma\check\phi'})\right]\notag
			\\
			&=\ \left(\rmd_\Sigma^\dagger\left(\frac{\rme^{\ad_{\check\phi'}}-\id}{\ad_{\check\phi'}}\right)\left(\frac{\id-\rme^{-\ad_{\check\phi'}}}{\ad_{\check\phi'}}\right)\right)(\rmd_\Sigma\check\phi')\notag
			\\
			&\kern1cm-\int_0^1\rmd t\,\star\rme^{t\ad_{\check\phi'}}\left[\left(\frac{\id-\rme^{-t\ad_{\check\phi'}}}{\ad_{\check\phi'}}\right)(\rmd_\Sigma\check\phi'),\left(\frac{\id-\rme^{-\ad_{\check\phi'}}}{\ad_{\check\phi'}}\right)({\star\rmd_\Sigma\check\phi'})\right]\notag
			\\
			&=\ (\Phi\circ\sfd^{\rm PCM}_{\rm CE})(\check\phi^+)~,\notag
		\end{align}
		where in the fourth step we have remembered the derivative formula $\dder[\rme^{\alpha(x)}]{x}=\int_0^1\rmd s\,\rme^{(1-s)\alpha(x)}\dder[\alpha(x)]{x}\rme^{s\alpha(x)}$ and in the last step~\eqref{eq:CEPCMRewritten}.

		\subsection{Quasi-isomorphism of Chevalley--Eilenberg algebras}

		In the previous section, we have found an isomorphism between the Chevalley--Eilenberg algebras $\sfA^{\rm PCM}$ and $\sfA'^{\rm PCM}$ for the principal chiral model. Using this, we shall now construct a quasi-isomorphism between the Chevalley--Eilenberg algebra $\sfA^{\rm shCS}$ of semi-holomorphic Chern--Simons theory and $\sfA^{\rm PCM}$.

		\paragraph{Morphism between $\sfA^{\rm shCS}$ and $\sfA'^{\rm PCM}$.}
		Firstly, following our discussion from \cref{sec:LinftyStructure}, we denote the dual fields and anti-fields of semi-holomorphic Chern--Simons theory by $\check c$, $\check A$, $\check A^+$, and $\check c^+$, respectively. The corresponding Chevalley--Eilenberg differential is then defined by
		\begin{equation}
			\begin{aligned}
				\sfd_{\rm CE}^{\rm shCS}\check c\ &\coloneqq\ -\tfrac12[\check c,\check c]~,
				\\
				\sfd_{\rm CE}^{\rm shCS}\check A\ &\coloneqq\ \bar\partial'_\rmF\check c+[\check A,\check c]~,
				\\
				\sfd_{\rm CE}^{\rm shCS}\check A^+\ &\coloneqq\ -\bar\partial'_\rmF\check A-\tfrac12[\check A,\check A]-[\check c,\check A^+]~,
				\\
				\sfd_{\rm CE}^{\rm shCS}\check c^+\ &\coloneqq\ \bar\partial'_\rmF\check A^++[\check A,\check A^+]-[\check c,\check c^+]~.
			\end{aligned}
		\end{equation}

		Next, we fix a smooth function $f$ on $\IC P^1$ which is equal $1$ in a neighbourhood around $z=0$ and $0$ in a neighbourhood around $z=\infty$, respectively. This is always possible by making use of a smooth partition of unity. We then define
		\begin{subequations}\label{eq:defMorphismSHCSPCM}
			\begin{equation}
				\begin{aligned}
					\Psi\,:\,\sfA^{\rm shCS}\ &\rightarrow\ \sfA'^{\rm PCM}~,
					\\
					\check c\ &\mapsto\ 0~,
					\\
					\check A\ &\mapsto\ \rme^{f\check\phi'}\check J'\rme^{-f\check\phi'}+\rme^{f\check\phi'}\bar\partial'_\rmF\rme^{-f\check\phi'}
					\\
					&\kern1cm =\ \sum_{n=0}^\infty\frac{1}{n!}\ad^n_{f\check\phi'}(\check J')-\sum_{n=0}^\infty\frac{1}{(n+1)!}\ad^n_{f\check\phi'}(\bar\partial'_\rmF(f\check\phi'))~,
					\\
					\check A^+\ &\mapsto\ \frac{z}{1-z^2}\star\rme^{f\check\phi'}\check\phi'^+\rme^{-f\check\phi'}\ =\ \frac{z}{1-z^2}\sum_{n=0}^\infty\frac{1}{n!}\ad^n_{f\check\phi'}({\star\check\phi'^+})~,
					\\
					\check c^+\ &\mapsto\ 0~,
				\end{aligned}
			\end{equation}
			where
			\begin{equation}
				\check J'\ \coloneqq\ \frac{1}{1-z^2}\check j'+\frac{z}{1-z^2}{\star\check j'}
			\end{equation}
		\end{subequations}
		and with $\check j'$ as defined in~\eqref{eq:CEPCMHat} (see also~\eqref{eq:solSemiHoloCS}); $\check J'$ is the Lax connection from~\eqref{eq:reducedSolSemiHoloCS}. This is indeed a morphism of graded algebras; in particular, it is not too difficult to see that $\Psi$ respects the boundary/pole conditions~\eqref{eq:shCSBDYConditions}.

		Before showing that the map defined above respects the differentials, we explain the notation used. The transition between the $L_\infty$-algebra picture and the differential-graded algebra picture is only well-defined under certain situations, for example, if a finite set of generators spans the underlying vector space of the $L_\infty$-algebra. This is not the case in our situation, however, as everything is local, we can focus on each point of $\Sigma \times \IC P^1$ separately. For example, consider the field $A\in \frL^{\rm shCS}_1$. Around each point $(z_0,x_0)$, we may expand any $A$ as $A(x,z)=\tau_\alpha\otimes\rmd x^\mu\,A^\alpha_\mu(x,z)+\tau_\alpha\otimes\rmd\bar z\,A^\alpha_{\bar z}(x,z)$, where $\tau_\alpha$ is a set of generators for the Lie algebra $\frg$. We assume that the set of linear functionals $(\frL^{\rm shCS})^*$ is generated by $\check A^\alpha_\mu(x,z)$ and $\check A^\alpha_{\bar z}(x,z)$ and their derivatives, which are defined by $\check A^\alpha_\mu(x,z)(A)\coloneqq A^\alpha_\mu(x,z)$ and $\check A^\alpha_{\bar z}(x,z)(A)\coloneqq A^\alpha_{\bar z}(x,z)$. The map $\Psi$ should be seen as a map of generators. The detail can be extracted by expanding both sides using a set of generators. For example, we have
		\begin{equation}
			\begin{gathered}
				\Psi(\check A^\alpha_{\bar z}(x,z))\ =\ -f_{\bar z}(z)\check\phi^\alpha(x)
				\ewith
				\bar\partial'_{\IC P^1}f(z)\ =\ \rmd\bar zf_{\bar z}(z)~.
			\end{gathered}
		\end{equation}

		Next, we show that
		\begin{equation}
			\sfd'^{\rm PCM}_{\rm CE}\circ\Psi\ =\ \Psi\circ\sfd^{\rm shCS}_{\rm CE}~,
		\end{equation}
		that is, $\Psi$ is a morphism of differential graded algebras. The only non-trivial parts to show are the action on $\check A^+$ and $\check c^+$. To calculate this, we shall make use of the identities
		\begin{subequations}
			\begin{equation}\label{eq:covcurvature}
				\begin{gathered}
					\bar\partial'_\rmF\check J'_{f\check\phi'}+\tfrac12[\check J'_{f\check\phi'},\check J'_{f\check\phi'}]\ =\ \rme^{f\check\phi'}\big(\bar\partial'_\rmF+\tfrac12[\check J',\check J']\big)\rme^{-f\check\phi'}~,
					\\
					\check J'_{f\check\phi'}\ \coloneqq\ \rme^{f\check\phi'}\check J'\rme^{-f\check\phi'}+\rme^{f\check\phi'}\bar\partial'_\rmF\rme^{-f\check\phi'}~.
				\end{gathered}
			\end{equation}
			and
			\begin{equation}
				\bar\partial'_\rmF\check J'+\tfrac12[\check J',\check J']\ =\ \frac{z}{1-z^2}\rmd_\Sigma{\star\check j'}
			\end{equation}
		\end{subequations}
		which are easily checked.\footnote{Note that $\frac{1}{1-z^2}$ and $\frac{z}{1-z^2}$ are in the kernel of $\bar\partial'_\rmF$.} Indeed, these identities immediately imply that
		\begin{equation}
			\begin{aligned}
				(\Psi\circ\sfd^{\rm shCS}_{\rm CE})(\check A^+)\ &=\ -\bar\partial'_\rmF\check J'_{f\check\phi'}-\tfrac12[\check J'_{f\check\phi'},\check J'_{f\check\phi'}]
				\\
				&=\ -\rme^{f\check\phi'}\big(\bar\partial'_\rmF \check J'+\tfrac12[\check J',\check J']\big)\rme^{-f\check\phi'}
				\\
				&=\ -\rme^{f\check\phi'}\left(\frac{z}{1-z^2}\rmd_\Sigma{\star\check j'}\right)\rme^{-f\check\phi'}
				\\
				&=\ (\sfd'^{\rm PCM}_{\rm CE}\circ\Psi)(\check A^+)~.
			\end{aligned}
		\end{equation}

		As for $\check c^+$, we need to show that $(\Psi\circ\sfd_{\rm CE}^{\rm shCS})(\check c^+)=0$ which is equivalent to proving the identity $\bar\partial'_\rmF\Psi(\check A^+)=-[\Psi(\check A),\Psi(\check A^+)]$. We have
		\begin{equation}
			\begin{aligned}
				\bar\partial'_\rmF\Psi(\check A^+)\ &= \ \bar\partial'_{\IC P^1}\left(\frac{z}{1-z^2}\star\rme^{f\check\phi'}\check\phi'^+\rme^{-f\check\phi'}\right)
				\\
				&=\ -\frac{z}{1-z^2}\star[\rme^{f\check\phi'}\bar\partial'_{\IC P^1}\rme^{-f\check\phi'},\rme^{f\check\phi'}\check\phi'^+\rme^{-f\check\phi'}]
				\\
				&=\ -[\Psi(\check A)|_{\IC P^1},\Psi(\check A^+)]
				\\
				&=\ -[\Psi(\check A),\Psi(\check A^+)]~.
			\end{aligned}
		\end{equation}

		\paragraph{Cohomology and quasi-isomorphism property.}
		The next step now is to verify that the morphism~\eqref{eq:defMorphismSHCSPCM} induces an isomorphism between cohomologies $H^\bullet_{\mu_1^{\rm shCS}}$ and $H^\bullet_{\mu_1^{\rm PCM}}$ of the complexes~\eqref{eq:shCSComplex} and~\eqref{eq:PCMComplex} underlying semi-holomorphic Chern--Simons theory and the principal chiral model, respectively. The argument here can be easily extended to define a contracting homotopy between the complexes.

		Firstly, given that $c$ is assumed to be smooth everywhere, see~\eqref{eq:shCSBDYConditions}, and since in this case the regularised and bare anti-holomorphic derivatives, see~\eqref{eq:regularisedAntiHoloDer}, agree, we can use the compactness of $\IC P^1$ to conclude that the only solution to $\bar\partial_\rmF c=0$ is $c$ being constant on $\Sigma\times\IC P^1$. However, since we assume that $c$ vanishes at $z=0$, this means that $c$ vanishes everywhere and thus, the zeroth cohomology group $H^0_{\mu_1^{\rm shCS}}=0$ vanishes. Since we have the non-degenerate pairing~\eqref{eq:shCSCyclicStructure}, we obtain an abstract Hodge--Kodaira decomposition for the complex\footnote{See~e.g.~\cite[Appendix B]{Jurco:2018sby} or~\cite[Appendix A]{Borsten:2024cfx} for reviews.} compatible with this pairing, which, in turn, descends to a non-degenerate pairing on cohomologies. Therefore, $H^3_{\mu_1^{\rm shCS}}\cong H^0_{\mu_1^{\rm shCS}}=0$ and so, the third cohomology group vanishes as well. This then also establishes the isomorphism $H^1_{\mu_1^{\rm shCS}}\cong H^2_{\mu_1^{\rm shCS}}$. To establish the quasi-isomorphism property of the composition
		\begin{equation}\label{eq:defE}
			\sfE\ \coloneqq\ \Phi^{-1}\circ\Psi\,:\,\sfA^{\rm shCS}\ \rightarrow\ \sfA^{\rm PCM}
		\end{equation}
		given by~\eqref{eq:PCMPCMPrimeMorphism} and~\eqref{eq:defMorphismSHCSPCM}, we now only need to show that it induces isomorphisms $H^p_{\mu_1^{\rm PCM}}\cong H^p_{\mu_1^{\rm shCS}}$ for $p=1,2$.

		To this end, we are switching back to the $L_\infty$-language. It is then not too difficult to see that at the level of complexes, $\sfE$ induces the morphism
		\begin{equation}\label{eq:morphismComplexes}
			\begin{aligned}
				\sfe_1\,:\,\sfCh(\frL^{\rm PCM})\ &\rightarrow\ \sfCh(\frL^{\rm shCS})~,
				\\
				\phi\ &\mapsto\
				\begin{pmatrix}
					A_\Sigma\\ A_{\IC P^1}
				\end{pmatrix}
				\ \coloneqq\
				\begin{pmatrix}
					\frac{1}{1-z^2}\rmd_\Sigma\phi+\frac{z}{1-z^2}{\star\rmd_\Sigma\phi}-f\rmd_\Sigma\phi
					\\
					-(\bar\partial'_{\IC P^1} f)\phi
				\end{pmatrix},
				\\
				\phi^+\ &\mapsto\
				\begin{pmatrix}
					A^+_\Sigma\\ A^+_{\IC P^1}
				\end{pmatrix}
				\ \coloneqq\
				\begin{pmatrix}
					0
					\\
					\frac{z}{1-z^2}{\star\phi^+}
				\end{pmatrix}.
			\end{aligned}
		\end{equation}
		We shall now establish that this induces an isomorphism on the underlying cohomology groups.

		We shall start with injectivity. Consider $\sfe_1(\phi)=\mu_1^{\rm shCS}(c)$, then~\eqref{eq:morphismComplexes} says that $\bar\partial'_{\IC P^1}(c+f\phi)=0$ and since $c+f\phi$ is a smooth function on $\IC P^1$, this is equivalent to $\bar\partial_{\IC P^1}(c+f\phi)=0$. Using the compactness of $\IC P^1$, we conclude that $c+f\phi=\gamma$ for some smooth function $\gamma$ on $\Sigma$. Using the boundary conditions~\eqref{eq:shCSBDYConditions} of $c$ and also those of $f$, it follows that $\phi=0$. Likewise, consider $\sfe_1(\phi^+)=\mu_1^{\rm shCS}(A_{\Sigma}+A_{\IC P^1})$. Then,~\eqref{eq:morphismComplexes} says that $\bar\partial'_{\IC P^1}A_\Sigma+\rmd_\Sigma A_{\IC P^1}=0$. Since $A_{\IC P^1}$ is smooth and since the Dolbeault cohomology group $H^{0,1}(\IC P^1)$ vanishes, we can write $A_{\IC P^1}=-\bar\partial_{\IC P^1}\gamma=-\bar\partial'_{\IC P^1}\gamma$ for some smooth function $\gamma$ on $\Sigma\times \IC P^1$ with $\gamma|_{z=\infty}=0$. Hence, $\bar\partial'_{\IC P^1}(A_\Sigma+\rmd_\Sigma\gamma)=0$ and so, the most general solution compatible with the boundary/pole conditions~\eqref{eq:shCSBDYConditions} is
		\begin{equation}\label{eq:generalSolutionASigma}
			A_\Sigma\ =\ \frac{1}{1-z^2}\rmd_\Sigma\gamma_0+\frac{z}{1-z^2}{\star\rmd_\Sigma\gamma_0}-\rmd_\Sigma\gamma
			\ewith
			\gamma_0\ \coloneqq\ \gamma|_{z=0}~.
		\end{equation}
		Consequently, $\rmd_\Sigma A_\Sigma=\frac{z}{1-z^2}\rmd_\Sigma{\star\rmd_\Sigma\gamma_0}=\frac{z}{1-z^2}{\star\phi^+}$ where the latter equality follows from the definition~\eqref{eq:morphismComplexes}. In turn, this implies that $\phi^+=-\rmd^\dagger_\Sigma\rmd_\Sigma\gamma_0=\mu_1^{\rm PCM}(\gamma_0)$ which vanishes in cohomology. In conclusion, we have shown that the morphism~\eqref{eq:morphismComplexes} is indeed injective when restricted to cohomology.

		Finally, we show surjectivity. To show that~\eqref{eq:morphismComplexes} restricts to a surjection at degree 1 in cohomology, we need to show that if $\mu_1^{\rm shCS}(A_\Sigma+A_{\IC P^1})=0$ then $A_\Sigma+A_{\IC P^1}$ is given by $\sfe_1(\phi)$ for some $\phi$ satisfying $\mu_1^{\rm PCM}(\phi)=0$. Suppose that $\mu_1^{\rm shCS}(A_{\Sigma}+A_{\IC P^1})=0$, then we have $\bar\partial'_{\IC P^1}A_\Sigma+\rmd_\Sigma A_{\IC P^1}=0$ and $\rmd_\Sigma A_{\Sigma}= 0$. By the same arguments as above, we have $A_{\IC P^1}=-\bar\partial'_{\IC P^1}\gamma$ for some smooth $\gamma$ on $\Sigma\times \IC P^1$. Therefore, the most general solution is again given by~\eqref{eq:generalSolutionASigma}. Since $\rmd_\Sigma A_{\Sigma}= 0$, we now have, $\mu_1^{\rm PCM}(\gamma_0)=0$. To see that $A_\Sigma+A_{\IC P^1}$ is given by $\sfe_1(\gamma_0)$, we note that $-\bar\partial'_{\IC P^1}(\gamma-f\gamma_0)=A_{\IC P^1}+(\bar\partial_{\IC P^1}f)\gamma_0$ and $\gamma-f\gamma_0$ satisfies the boundary/pole conditions~\eqref{eq:shCSBDYConditions}, so $\sfe_1(\gamma_0)$ and $A_{\IC P^1}$ are gauge equivalent. To show that~\eqref{eq:morphismComplexes} restricts to a surjection at degree 2 in cohomology, we need to show that if $\mu_1^{\rm shCS}(A^+_\Sigma+A^+_{\IC P^1})=0$ then $A^+_\Sigma+A^+_{\IC P^1}$ is given by $\sfe_1(\phi^+)$ for some $\phi^+$. Without loss of generality, $A_\Sigma^+=0$.\footnote{This follows directly from the fact that $\bar\partial'_{\IC P^1}:\frL^{\rm shCS}_{1,\,\Sigma}\rightarrow\frL^{\rm shCS}_{2,\,\Sigma}$ is invertible. To see this latter fact, we first note that $\bar\partial_{\IC P^1}:\Omega^{0,0}(\IC P^1)\rightarrow\Omega^{0,1}(\IC P^1)$ has a section $\rho:\Omega^{0,1}(\IC P^1)\rightarrow\Omega^{0,0}(\IC P^1)$ with $\rho(\alpha)(z=0)=0$ given by $\rho(\alpha)(\lambda)=-\frac{1}{2\pi\rmi}\int_{\IC P^1}[\rmd \lambda^\prime\lambda^\prime]\frac{[\xi\lambda]}{[\xi\lambda'][\lambda\lambda^\prime]}\wedge\alpha(\lambda^\prime)$ with $\lambda_{\dot\alpha}$ homogeneous coordinates on $\IC P^1$, $[\lambda'\lambda]\coloneqq\varepsilon^{\dot\alpha\dot\beta}\lambda'_{\dot\alpha}\lambda_{\dot\beta}$, and $\xi\coloneqq(1,0)$; see~\cite[Eq.~6.7]{Wolf:2010av}. Furthermore, as $\rho$ is given by an integral on $\IC P^1$, i.e.~a compact space, it maps any smooth family of one-forms to a smooth family of functions.

		Using this, we can now see that $\bar\partial'_{\IC P^1}:\frL^{\rm shCS}_{1,\,\Sigma}\rightarrow\frL^{\rm shCS}_{2,\,\Sigma}$ is an isomorphism. In particular, decompose $\frL^{\rm shCS}_{1,\,\Sigma}$ into self-dual and anti-self-dual parts on $\Sigma$, and each part is proved similarly. So without loss of generality, we can restrict ourselves to the self-dual part only. Next, the kernel of $\bar\partial'_{\IC P^1}:\frL^{\rm shCS}_{1,\,\Sigma}\rightarrow\frL^{\rm shCS}_{2,\,\Sigma}$ consists of smooth one-form families, depending on $\Sigma$, of meromorphic functions with simple poles at $z=1$. By the Mittag--Leffler expansion, the only such functions are of the form $\frac{1}{1-z}A_++A_0$ for any pair of self-dual one-forms $A_+$ and $A_0$ on $\Sigma$. However, there is no non-trivial element that satisfies the boundary conditions~\eqref{eq:shCSBDYConditions}. Hence, the kernel is trivial and so, $\bar\partial'_{\IC P^1}:\frL^{\rm shCS}_{1,\,\Sigma}\rightarrow\frL^{\rm shCS}_{2,\,\Sigma}$ is injective. To see surjectivity, for $A^+_{\Sigma,\,+}\in\frL^{\rm shCS}_{2,\,\Sigma}$, we have $A^+_{\Sigma,\,+}\big|_{U_+}=\frac{1}{1-z}A^+_{\Sigma,+,1}$ such that $A^+_{\Sigma,+,1}$ is smooth. Consider another open set $U_\infty\not\ni\{1\}$ such that $U_+\cup U_\infty=\IC P^1$ together with a smooth partition of unity $(\theta_+,\theta_\infty)$ subordinate to this cover. We can write $A^+_{\Sigma,+}=\frac{1}{1-z}A^+_{\Sigma,+,1}\theta_++A^+_{\Sigma,+}\theta_\infty$. Now, we have
		$$
			\bar\partial'_{\IC P^1}\left[\frac{1}{1-z} \rho(A^+_{\Sigma,+,1}\theta_+)+\rho(A^+_{\Sigma,+}\theta_\infty)+\frac{z}{1-z}\left(\rho(A^+_{\Sigma,+}\theta_\infty)(z=\infty)\right)\right]\ =\ A^+_{\Sigma,+}
		$$
		with $\rho$ as given above. Hence, $\bar\partial'_{\IC P^1}:\frL^{\rm shCS}_{1,\,\Sigma}\rightarrow\frL^{\rm shCS}_{2,\,\Sigma}$ is surjective and thus, an isomorphism.} Then, $\mu_1^{\rm shCS}(A^+_\Sigma+A^+_{\IC P^1})=0$ reduces to $\bar\partial'_{\IC P^1}A^+_{\IC P^1}=0$. It is not too difficult to see that the only solution compatible with the boundary/pole conditions~\eqref{eq:shCSBDYConditions} is $\frac{z}{1-z^2}\alpha$ with $\alpha$ a smooth two-form on $\Sigma$. Altogether, we have established surjectivity.

		\paragraph{Cyclicity.}
		Above we have demonstrated that the morphism~\eqref{eq:defE} is a quasi-isomorphism of differential graded algebras and thus, dually, of $L_\infty$-algebras. What remains to check that it also respects the inner products, that is, the cyclic structure in the sense of~\cite{Kajiura:2003ax}. In the $L_\infty$-algebra picture, the quasi-isomorphism~\eqref{eq:defE} is given by a set of graded anti-symmetric multilinear maps $\sfe_i:\frL^{\rm PCM}\times\cdots\times\frL^{\rm PCM}\rightarrow\frL^{\rm shCS}$ of degree $1-i$; see~\eqref{eq:morphismComplexes} for $\sfe_1$. In particular, we wish to verify the relations
		\begin{equation}\label{eq:defCyclicMorph}
			\begin{gathered}
				\inner{a_1}{a_2}^{\rm PCM}\ =\ \inner{\sfe_1(a_1)}{\sfe_1(a_2)}^{\rm shCS}~,
				\\
				\sum_{j+k=i}\sum_{\sigma\in\overline{\mathsf{Sh}}(j;i)}\chi(\sigma;a_1,\ldots,a_i)\inner{\sfe_j(a_{\sigma(1)},\ldots,a_{\sigma(j)})}{\sfe_k(a_{\sigma(j+1)},\ldots,a_{\sigma(i)})}^{\rm shCS}\ =\ 0
			 \end{gathered}
		\end{equation}
		for all $i\geq 3$ and for all $a_1,\ldots,a_i\in\frL^{\rm PCM}$. Above, $\overline{\mathsf{Sh}}(j;i)$ refers to the set of $(j;i)$-unshuffles, that is, permutations $\sigma$ of $\{1,\ldots,i\}$ such that $\sigma(1)<\sigma(2)<\cdots<\sigma(j)$ and $\sigma(j+1)<\sigma(j+2)<\cdots<\sigma(i)$, and $\chi(\sigma;a_1,\ldots,a_i)$ is the corresponding Koszul sign making the expression graded anti-symmetric.

		The first relation in~\eqref{eq:defCyclicMorph} is easily verified. Indeed, the only non-trivial case is when $a_1=\phi\in\frL^{\rm PCM}_1$ and $a_2=\phi^+\in\frL^{\rm PCM}_2$, that is, a pairing of a field and anti-field. Using~\eqref{eq:morphismComplexes}, we have
		\begin{equation}
			\begin{aligned}
				\inner{\sfe_1(\phi)}{\sfe_1(\phi^+)}^{\rm shCS}\ &=\ \frac{\rmi}{2\pi}\int_{\IC P^1\times\Sigma}\Omega\wedge\innerLarge{-(\bar\partial'_{\IC P^1}f)\phi}{\frac{z}{1-z^2}{\star\phi^+}}
				\\
				&=\ \frac{\rmi}{2\pi}\int_{\IC P^1\times\Sigma}\Omega\wedge\innerLarge{-(\bar\partial_{\IC P^1}f)\phi}{\frac{z}{1-z^2}{\star\phi^+}}
				\\
				&=\ f(0)\int_\Sigma\inner{\phi}{\star\phi^+}
				\\
				&=\ \inner{\phi}{\phi^+}^{\rm PCM}~,
			\end{aligned}
		\end{equation}
		where in the third step we have integrated by parts and used~\eqref{eq:deltaFunctionIdentities}, and in the last we have used $f(0)=1$.

		We now move on to the second relation in~\eqref{eq:defCyclicMorph}. As the expression is graded anti-symmetric, we can use the dual Chevalley--Eilenberg picture. Upon adding the first relation to the second, we thus need to verify
		\begin{equation}
			\inner{\sfE(\check A)}{\sfE(\check A^+)}^{\rm shCS}\ =\ \inner{\check\phi}{\check\phi^+}^{\rm PCM}
		\end{equation}
		to establish cyclicity. Using the explicit formulas~\eqref{eq:PCMPCMPrimeMorphism} and~\eqref{eq:defMorphismSHCSPCM}, we note that $\sfE(\check A^+)$ is a two-form on $\Sigma$ and so,
		\begin{equation}
			\begin{aligned}
				\inner{\sfE(\check A)}{\sfE(\check A^+)}^{\rm shCS}\ &=\ -\frac{\rmi}{2\pi}\int_{\IC P^1\times\Sigma}\Omega\wedge\innerLarge{\rme^{f\check\phi}\bar\partial'_{\IC P^1}\rme^{-f\check\phi}}{\frac{z}{1-z^2}\star\rme^{f\check\phi}\Phi^{-1}(\check\phi'^+)\rme^{-f\check\phi}}
				\\
				&=\ -\frac{\rmi}{2\pi}\int_{\IC P^1\times\Sigma}\Omega\wedge\innerLarge{\rme^{f\check\phi}\bar\partial_{\IC P^1}\rme^{-f\check\phi}}{\frac{z}{1-z^2}\star\rme^{f\check\phi}\Phi^{-1}(\check\phi'^+)\rme^{-f\check\phi}}
				\\
				&=\ \frac{\rmi}{2\pi}\int_{\IC P^1\times\Sigma}\Omega\wedge\innerLarge{(\bar\partial_{\IC P^1}f)\check\phi}{\frac{z}{1-z^2}\star\Phi^{-1}(\check\phi'^+)}
				\\
				&=\ \inner{\check\phi}{\Phi^{-1}(\check\phi'^+)}^{\rm PCM}
				\\
				&=\ \sum_{n=0}^\infty\frac{B_n}{n!}\inner{\check\phi}{\ad_{\check\phi}^n(\check\phi^+)}^{\rm PCM}
				\\
				&=\ \sum_{n=0}^\infty\frac{(-1)^nB_n}{n!}\inner{\ad_{\check\phi}^n(\check\phi)}{\check\phi^+}^{\rm PCM}
				\\
				&=\ \inner{\check\phi}{\check\phi^+}^{\rm PCM}~,
			\end{aligned}
		\end{equation}
		where in the third step we have used the invariance property of $\inner{-}{-}$ on $\frg$, in the fourth step we have integrated by parts and used~\eqref{eq:deltaFunctionIdentities} and $f(0)=1$, and in the fifth step we have again used the invariance of $\inner{-}{-}$ on $\frg$.




	\end{body}

\end{document}